\documentclass[]{amsart}
\usepackage{amsmath}
\usepackage{amsthm}
\usepackage{amssymb}
\usepackage{amscd}
\usepackage{amsfonts}
\usepackage{amsbsy}
\usepackage{epsfig,afterpage}

\usepackage[usenames,dvipsnames]{color}

\begin{document}

\title{The Singing Arc: The Oldest Memristor?}

\author[J.M. Ginoux, B. Rossetto]
{Jean-Marc Ginoux$^1$ and Bruno Rossetto$^2$}

\address{$^1$ Laboratoire {\sc Protee}, I.U.T. de Toulon, Universit\'{e} du Sud, BP 20132, F-83957 La Garde Cedex, France, ginoux@univ-tln.fr, http://ginoux.univ-tln.fr}
\email{ginoux@univ-tln.fr}

\address{$^2$ Laboratoire {\sc Protee}, I.U.T. de Toulon, Universit\'{e} du Sud, BP 20132, F-83957 La Garde Cedex, France, rossetto@univ-tln.fr}



\begin{abstract}
On April 30$^{th}$ 2008, the journal \textit{Nature} announced that the missing circuit element, postulated thirty-seven years before by Professor Leon O. Chua has been found. Thus, after the capacitor, the resistor and the inductor, the existence of a fourth fundamental element of electronic circuits called ``memristor'' was established. In order to point out the importance of such a discovery, the aim of this article is first to propose an overview of the manner with which the three others have been invented during the past centuries. Then, a comparison between the main properties of the singing arc, i.e. a forerunner device of the triode used in Wireless Telegraphy, and that of the memristor will enable to state that the singing arc could be considered as the oldest memristor.
\end{abstract}

\maketitle

\section{Introduction}

Contrary to what one might think, it is not by experimenting, but by logical deduction that Professor L. O. Chua was able to postulate the existence of a missing circuit element. In his now famous publication of 1971, \cite{Chua1971} considered the three basic building blocks of an electric circuit: the capacitor, the resistor and the inductor as well as the three laws linking the four fundamental circuit variables, namely, the electric \textit{current} $i$, the \textit{voltage} $v$, the \textit{charge} $q$ and the \textit{magnetic flux} $\varphi$. Then, \cite[p. 507]{Chua1971} explained that:\\

\begin{quote}

``By the \textit{axiomatic} definition of the three classical circuits elements, namely, the \textit{resistor} (defined by a relationship between $v$ and $i$), the \textit{inductor} (defined by a relationship between $\varphi$ and $i$), and the \textit{capacitor} (defined by a relationship between $q$ and $v$). Only one relationship remains undefined, the relationship between $\varphi$ and $q$.''

\end{quote}

\vspace{0.1in}

He thus concluded from the logical as well as axiomatic points of view, that it is necessary, for the sake of \textit{completeness}, to postulate the existence of a fourth circuit element to which he gave the name \textit{memristor} since it behaves like a nonlinear resistor with memory. The manner with which these three first laws that he calls: ``relationships'' have been discovered will be briefly recalled in Sec. 2 while the fourth law that he has established as well as the main properties of the memristor will be presented in Sec. 3. Then, it will be shown in Sec. 4 that the singing arc which exhibits analogous characteristic to the missing circuit element can be considered as the oldest memristor.

\section{Fundamental passive circuits elements}
\label{Fund}

Anyone who has taken a physics class is familiar with the capacitor, the resistor, the inductor and of course the generator. But the question that one can ask is when and how these devices have been invented. This section aims to answer this question.

\subsection{Le condensatore and Volta's law}

In a way we could say that the history of the \textit{capacitor} begins with the History of Physics, i.e. at the time of the Ancient Greece. Indeed, according to many historians of sciences, Thales of Miletus (624-547 B.C.) who is generally considered as the first physicist, has highlighted the fact that the rubbed \textit{amber} ($\eta\lambda\varepsilon\kappa\tau\varrho\o\nu$ elektron in Greek) has the property of attracting light objects, such as pieces of straw or fabric. At the beginning of the seventeenth century, the English physician, physicist and natural philosopher William Gilbert (1544-1603) extended in his book \textit{De Magnete, Magneticisque Corporibus, et de Magno Magnete Tellure} (On the Magnet and Magnetic Bodies, and on the Great Magnet the Earth) published in 1600, the number of substances capable of ``becoming electric'' (triboelectric), that is to say, after friction they exhibit the same property as amber. In the second chapter entitled: On the Magnetick Coition, and first on the Attraction of Amber, or more truly, on the Attaching of Bodies to Amber, \cite[p. 48]{Gilbert1600} wrote:\\

\begin{quote}
``\textit{Nam non solum succinum, \& gagates (vt illi putant) allectant corpuscula}.
For it is not only amber and jet (as they suppose) which entice small bodies; but Diamond, Sapphire, Carbuncle, Iris gem, Opal, Amethyst, Vincentina, and Bristolla (an English gem or spar), Beryl, and Crystal do the same.''
\end{quote}

\vspace{0.1in}

\begin{figure}[htbp]
\centerline{\includegraphics[width=10.16cm,height=6.35cm]{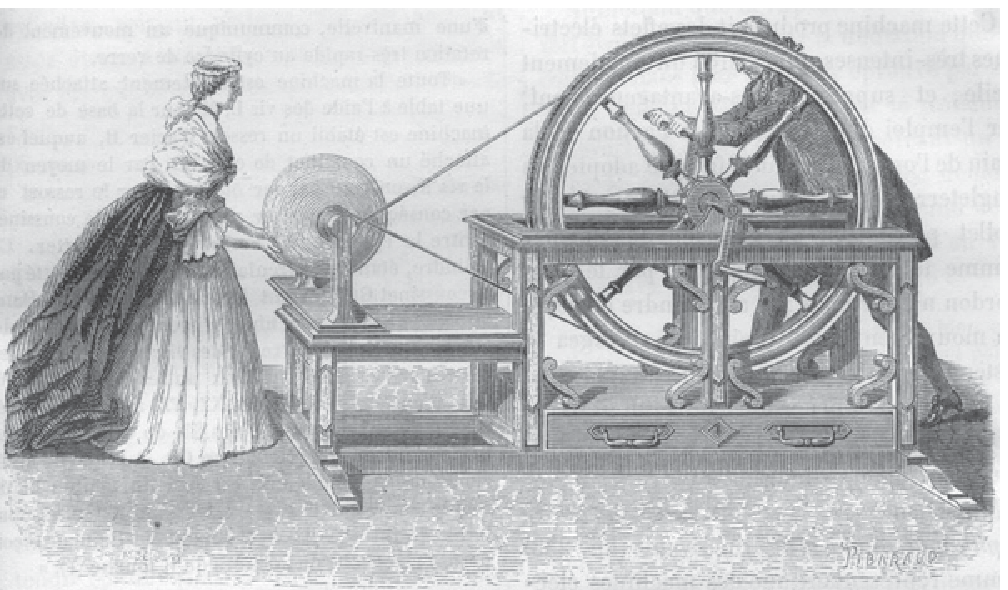}}
\caption{Abb\'{e} Nollet's friction machine, \cite[p. 24]{Nollet1753}.}
\label{fig1}
\end{figure}

This list has grown with the glasses of various compositions, such as resins, or sulfur. After the body is rubbed with dry hands, or with paper, leather or fabric, feathers and straw are more or less attracted. But the observed phenomena, did not received spectacular interests before the 1700s. During this period, the electrostatic generator, then called \textit{friction machine}, appeared. The German scientist Otto von Guericke (1602-1686) is considered as the inventor of a primitive form of frictional electrical machine. In 1671, he had the idea to make a sulphur globe that could be rotated and rubbed by hand. In 1707, Isaac Newton (1642-1727) suggested to use a glass globe instead of a sulphur one and in 1709 Francis Hauksbee (1666-1713) improved the basic design, by rotating a glass sphere to be rotated rapidly against a woollen cloth. Thus, in the middle of the eighteenth century, the \textit{Electricty} becomes fashionable. In France, it was widely popularized by the French clergyman and physicist Jean-Antoine Nollet (1700-1770) known as Abb\'{e} Nollet. On March 14$^{th}$ 1746, he made one of his most spectacular experiments in the galerie des Glaces at Versailles in the presence of his Majesty the King Louis XV. He electrified a hundred eighty royal guards holding hands with an electrostatic generator (See Fig. \ref{fig1}) positioned at the beginning of the human chain.

\newpage

The same year in January, Nollet who has been asked to translate a letter send by the Dutch scientist Pieter Musschenbroek (1692-1761), then professor at Leiden, to Ren\'{e} R\'{e}aumur (1683-1757) and written in Latin, had credited him erroneously for the discovery\footnote{This episode has been clarified by Edmond Hoppe in his History of Physics \cite[p. 448]{Hoppe1928}.} of a device that he called the ``Leiden jar''. In fact, this invention was made one year earlier, on October 10$^{th}$ 1745 by a prelate named Ewald Georg von Kleist (1700-1748) who wanted to confine electricity in a bottle. So, he filled the bottle with water closed by a cork pierced by a nail. Holding the bottle with one hand he attached the nail against an electrostatic generator. Then, after having moved away the bottle from the machine, he touched the nail with his other hand and received a violent electric discharge. In fact von Kleist had invented the first capacitor.

\begin{figure}[htbp]
\centerline{\includegraphics[width=5cm,height=6cm]{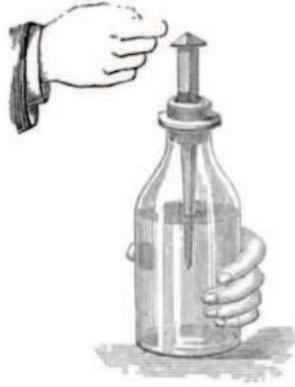}}
\caption{The Leyden jar.}
\label{fig2}
\end{figure}

However, it's the Italian physicist Alessandro Volta (1745-1827) who gave it this name half a century later. Volta is generally well-known for his invention in 1800 of the electric battery. Ten years before, in 1771, he had improved and popularized the electrophorus: a capacitive generator used to produce electrostatic charge via the process of electrostatic induction. Then, he published his research about the electrophorus in the \textit{Philosophical transactions of the Royal Society of London} in 1782 and they were reproduced in his Complete Works in 1816. In this paper \cite[vol. 1, part 2, p. 50]{Volta1816} wrote:\\

\begin{quote}
``Questo artificio, voi forse gi\`{a} l'indovinate, consiste a riunire all eletrometro medesimo il \textit{Condensatore}\footnote{``This device, you probably already guessed, is to bring together all the same eletrometers under the name \textit{Capacitor}.''}.''
\end{quote}

\vspace{0.1in}

Then, \cite[vol. 1, part 1, p. 250]{Volta1816} exposed the law that Prof. Chua called ``relationship'' linking the tension $v$, the charge $q$ and the capacity $C$ of the capacitor:

\begin{quote}
``Ci\`{o} che abbiam detto comprendersi facilmente che la tensione debbe essere in ragione inversa delle capacita, ci viene poi mostrato nella maniera più chiara dall esperienza\footnote{``What we have said can easily be understood that the voltage ought to be in inverse proportion to the capacity, as it is shown clearly by experiments.''}.''
\end{quote}

This law is generally expressed as:

\begin{equation}
q = Cv
\end{equation}

\vspace{0.1in}

We suggest to name this relationship: Volta's law. During the same period Benjamin Franklin (1706-1790) invented the terminology ``electric charge'' while Henry Cavendish (1731-1810) introduced the notion of electrostatic capacity of a sphere one inch in diameter that he called ``globular inches''.

\begin{figure}[htbp]
\centerline{\includegraphics[width=7.94cm,height=10.1cm]{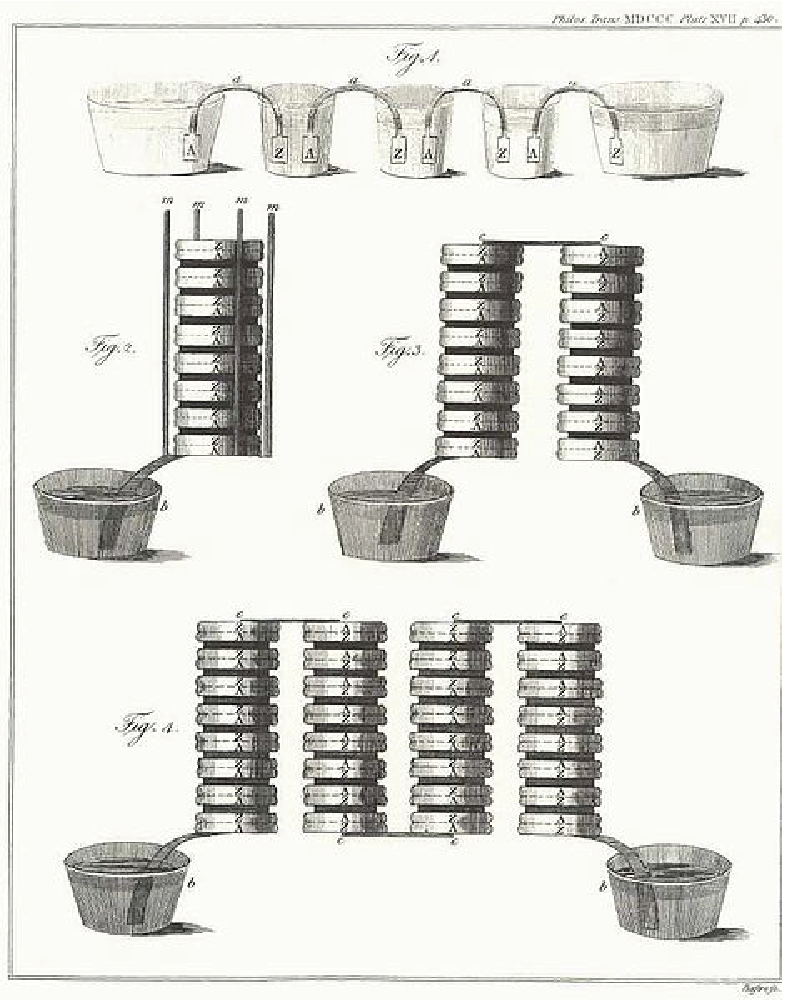}}
\caption{Voltaic pile, \cite[p. 430]{Volta1800}.}
\label{fig3}
\end{figure}

Unfortunately, this series of papers on electrical phenomena, which had appeared in the \textit{Philosophical transactions of the Royal Society of London}, i.e. the bulk of his electrical experiments, did not become known until they were collected and published by James Clerk Maxwell (1831-1879) a century later, in 1879, long after other scientists had been credited with the same results.
On March 20$^{th}$ 1800, Volta wrote his first letter to Joseph Banks (1743-1820) in which he described the first true battery which came to be known as the Voltaic Pile (See Fig. \ref{fig3}). This letter was send for publication to the \textit{Philosophical transactions} and read at the \textit{Royal Society of London} on June 26$^{th}$ 1800 (See \cite{Volta1800}).

On November 7$^{th}$ and 20$^{th}$ 1801, Volta presented his device at the Institut de France before Napol\'{e}on who awarded him a gold medal. Although Volta made use of the term ``tensione'' in his articles introducing thus in a qualitative way the \textit{voltage}, it's the French Physicist and Mathematician Andr\'{e} Marie Amp\`{e}re (1775-1836) who defined accurately the concepts of \textit{current} and \textit{voltage} in his first m\'{e}moir of 1820 \cite[p. 59]{Amp1820} wrote:

\begin{quote}
``L'action \'{e}lectromotrice se manifeste par deux sortes d'effets que je crois devoir d'abord distinguer par
une d\'{e}finition pr\'{e}cise. J'appellerai le premier tension \'{e}lectrique, le second courant \'{e}lectrique\footnote{``The electromotive action is characterized by two kinds of effects that I believe I must first distinguish by a precise definition. I will call the first voltage, the second electric current.''}.''
\end{quote}

\vspace{0.1in}

Nevertheless, it was unknown at that time among most of the physicists, including Amp\`{e}re, the precise idea of \textit{resistance} and of a possible ``relationship'' between the \textit{voltage} of a voltaic pile and the \textit{current} it produces in a conductor.
For further research on Alessandro Volta see for example \cite{Pancaldi2003}.

\subsection{Elektrischer Widerstand and Ohm's law}

The German physicist Georg Simon Ohm (1789-1854) started his experiments on electrical currents in 1825 while using a voltaic pile. The year after, he replaced it by copper-bismuth thermoelectric couples and could thus established his famous law \cite{Ohm1826a,Ohm1826b}:

\begin{equation}
v = Ri
\end{equation}

\vspace{0.1in}

Ohm defined thus the concept of ``elektrischer Widerstand'' (resistance) and of resistivity with great precision. Their measurements, in relative value, became thus accessible. In 1827, he summarized his research in a book entitled: \textit{Die galvanische Kette, mathematisch bearbeitet (The galvanic circuit invesitgated mathematically)} \cite{Ohm1827} in which he introduced the concept of potential difference (P.D.) that he called ``gef\"{a}lle''. \cite[p. 19]{Ohm1827} wrote:\\

\begin{quote}
``... wo man unter dem Ausdrucke "Gef\"{a}lle" die Differenz solcher Ordinaten zu verstehen hat, die zu zwei um die L\"{a}ngeneinheit von einander entfernten Stellen geh\"{o}ren\footnote{``That is to say, the steepness or extent of the fall of potential in equal lengths of different portions of the circuit of like material but of different size, will be directly proportional to the resistance of such portions.''}''.

\end{quote}

\vspace{0.1in}

For further research on Georg Ohm see for example \cite{Pourprix1989, Pourprix1995}.

\subsection{Induction and Faraday's law}

\begin{figure}[htbp]
\centerline{\includegraphics[width=9.2cm,height=5.5cm]{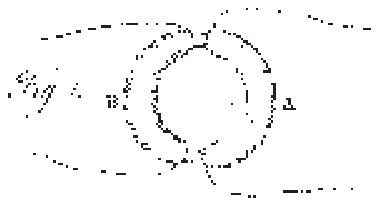}}
\caption{Faraday's iron-ring coil apparatus, \cite[p. 131]{Faraday1832}.}
\label{fig4}
\end{figure}

Everybody knows the story of Michael Faraday (1791-1867) apprentice bookbinder who attended lectures by the eminent English chemist Humphry Davy (1778-1829) of the Royal Institution. When Davy damaged his eyesight in an accident with nitrogen trichloride he hired Faraday as secretary and then as Chemical assistant. After graduating Faraday became his successor in 1827 and later one of the most famous physicists. On August 29$^{th}$ 1831, Faraday made his first experimental demonstration of electromagnetic induction. He wrapped two wires around opposite sides of an iron-ring coil (See Fig. \ref{fig4}). In one side (A) he sent a current by connecting it to a battery while the opposite side (B) was attached to a galvanometer. At the moment of opening and closing the current, the galvanometer indicated current of opposite sense. He called the first coil an ``inductor'' with a primary current passing through it. He called the current passing through the second coil and galvanometer the secondary current.

The phenomenon he called ``induction'' was caused by the change in magnetic flux that occurred when the battery was connected, or disconnected. In 1834, Heinrich Lenz (1804-1865) formulated his famous law which describes ``flux through the circuit'' and gave the direction of the induced electromotive force and current resulting from electromagnetic induction that \cite[p. 125, p. 129, p. 139]{Faraday1832} had described thus:

\begin{quote}
``The power which electricity of tension possesses of causing an opposite electrical state in its vicinity has been expressed by the general term Induction; which, as it has been received into scientific language, may also, with propriety, be used in the same general sense to express the power which electrical currents may possess of inducing any particular state upon matter in their immediate neighborhood, otherwise indifferent. It is with this meaning that I purpose using it in the present paper. [...] For the purpose of avoiding periphrasis, I propose to call this action of the current from the voltaic battery, \textit{volta-electric} induction. [...] but, as a distinction in language is still necessary, I propose to call the agency thus exerted by ordinary magnets, \textit{magneto-electric} or \textit{magnelectric} induction.''
\end{quote}

\vspace{0.1in}

Ten years later, starting from the work of Faraday and Lenz, Franz Neumann (1798-1895) established the mathematical law of induction. Based on the idea suggested by the rule of Lenz that the induced currents are created in the case of the motion, by working against electromagnetic forces, he derived the mathematical law for this phenomenon in the form:

\begin{equation}
e = -\frac{d\varphi}{dt}
\end{equation}

\vspace{0.1in}

often called ``Faraday's law'', although the works of Faraday did not contain any mathematical formula. This law means that the induced electromotive force is proportional to the ``magnetic flux'' $\varphi$ cut per unit time, or the change per unit time of the induced ``magnetic flux'' $\varphi$ embraced by the circuit. The self-inductance $L$ of a coil traversed by a changing current of intensity $i$ is then related to the ``magnetic flux'' $\varphi$ by the formula:

\begin{equation}
\varphi = Li
\end{equation}

\vspace{0.1in}

For further research on Michael Faraday see for example S. \cite{Thompson1901}.\\

As recalled in the previous sections, for each of the three circuit elements: capacitor ($C$), resistor ($R$) and inductor ($L$), a \textit{linear} relationship, i.e. a \textit{linear} law linking amongst themselves, a different pair of variables $(q, v)$, $(v, i)$ and $(\varphi, i)$, has been discovered. These three linear laws seem to suggest that circuit theory was dominated by a kind of ``linear thinking''.

\section{The singing Arc}

In fact, in the late nineteenth century, a French engineer Jean-Marie-Anatole G\'{e}rard-Lescuyer observed in an electromagnetic device, a nonlinear behavior that he considered then as paradoxical. By coupling a dynamo acting as a generator to a magneto-electric machine he built what was called a series dynamo machine. \cite[p. 215]{GL2} noticed a surprising phenomenon which he described as follows.\\

\begin{quote}
``As soon as the circuit is closed the magnetoelectrical machine begins to
move; it tends to take a regulated velocity in accordance with the intensity of
the current by which it is excited; but suddenly it slackens its speed, stops,
and starts again in the opposite direction, to stop again and rotate in the same
direction as before. In a word, it receives a regular reciprocating motion
which lasts as long as the current that produces it.''
\end{quote}

\vspace{0.1in}

He thus reported periodic reversals in the rotation of the magneto-electric machine although the source current was constant. He considered this
phenomenon as an ``electrodynamical paradox''. According to him, an increase in the velocity of the magneto-electric machine gives rise to a current in the opposite direction which reverses the polarity of the inductors and, then reverses its rotation. Many years later, the Dutch physicist Baltahzar Van der Pol (1889-1959) will establish that this nonlinear phenomenon belongs to the class of ``relaxation oscillations'' in the second Dutch version \cite[p. 39]{VdP1926b} of his eponymous article on ``Relaxation-Oscillations'' \cite{VdP1926d}. It has been pointed out by one of us \cite{Gith} that there was at least four versions of this article:\\

\begin{itemize}
\item[(i)] Over Relaxatietrillingen, \textit{Physica}, 6, 154--157,\\ \cite{VdP1926a},
\item[(ii)] Over Relaxatie-trillingen, \textit{Tijdschrift Nederlandsch Radiogenoot} 3, 25--40, \cite{VdP1926b},
\item[(iii)] \"{U}ber Relaxationsschwingungen, \textit{Jb. Drahtl. Telegr.} 28, 178--184,\\ \cite{VdP1926c},
\item[(iv)] On relaxation-oscillations, \textit{Philosophical Magazine}, serie 7, (2) 978--992, \cite{VdP1926d}.
\end{itemize}

\vspace{0.1in}

The English version \cite{VdP1926d}, the best known of his article, was the last to be published in the last two months of 1926. Van der Pol gave the explanation of G\'{e}rard-Lescuyer's electrodynamical paradox by showing that the relation between the potential difference $v$ across the dynamo's terminals and the current intensity $i$ through it may be represented by a nonlinear function. In other words, as we will see further, the tension-current characteristic ($v,i$) of the dynamo can be modeled by a nonlinear function.\\

\begin{figure}[htbp]
\centerline{\includegraphics[width=7cm,height=4.5cm]{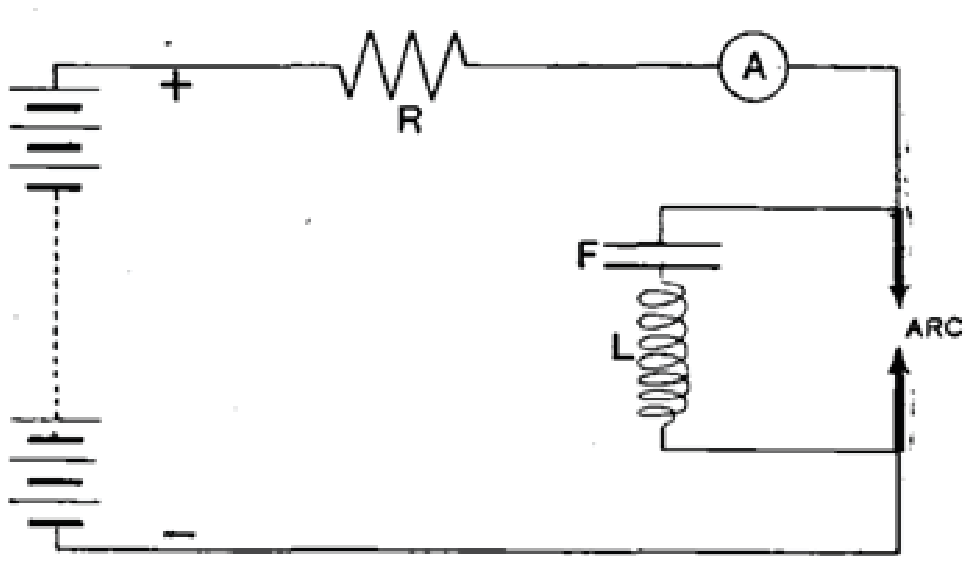}}
\caption{Singing arc circuit diagram, \cite[p. 248]{Dud1900a}}
\label{fig5}
\end{figure}

At the end of the nineteenth century a device, ancestor of the incandescent lamp, called electric arc was used for illumination of lighthouses and cities\footnote{The electrical arc (artificial in contrast to the flash of lightning) is associated with the electrical discharge produced between the ends of two electrodes (e.g. carbon), which also emits light. It is still used today in cinema projectors, plasma and thermal metallurgy in ``arc welding'' or smelting (arc furnaces).}. It exhibited independently of its low intensity light, a major drawback: the noise generated by the electrical discharge disturbed the residents. In London, the British physicist William Du Bois Duddell (1872-1917), was commissioned in 1899 by the British authorities to solve this problem. He had the idea of combining an oscillatory circuit composed of an inductor $L$ and a capacitor of capacitance $C$ ($F$ in Fig. \ref{fig5}) with the electric arc and then stopped the rustling (see Fig. \ref{fig5}).

Following his investigations, \cite[p. 248]{Dud1900a} reported in an article entitled \textit{Musical Arc}:\\

\begin{quote}
``A direct-current arc of suitable length and current between solid carbons, will give out a musical note if it be shunted with a condenser in series with a self-induction, as in Fig. \ref{fig5} ...''
\end{quote}

\vspace{0.1in}

In fact, Duddell had designed an oscillatory circuit susceptible to produce sounds and more than that: electromagnetic waves. Thus, this apparatus was used as emitter and receiver for the wireless telegraphy till the advent of the triode. By producing spark, the \textit{musical arc}, or \textit{singing arc}, or \textit{Duddell's arc}, generated electromagnetic waves highlighted by the experiments of \cite{Hertz}. In France, the engineer Andr\'{e} Blondel (1863-1938) who was assigned to the Lighthouses and Beacons Service, to conduct research on the electric arc in order to improve the outdated French electrical system, made an extensive analysis of the musical arc \cite{Blo1905} using Duddell's circuit (Fig. \ref{fig5}). He presented his main contribution on this subject in a fundamental paper entitled: ``On the singing arc phenomenon'' in which he pointed out the fact that the arc oscillations were ``nearly sinusoidal'' \cite[p. 46]{Blo1905}. Although the ``audion'' was invented in 1907 by Lee de Forest (1873-1961) \cite{For07}, it was not until the First World War that it was widely distributed for military and commercial purposes under the name of ``triode'' coined by the British physicist William Eccles (1875-1966) (\cite[p. 11]{Eccles1919}). In April 1919, the French engineer Paul Janet (1863-1937) exhibited an analogy between the \textit{series dynamo machine}, the \textit{singing arc} and the \textit{triode}. \cite[p. 764]{Janet} wrote about G\'{e}rard-Lescuyer's experiment with the series dynamo machine:

\begin{quote}

``It seemed interesting to me to report unexpected analogies between this
experiment and the sustained oscillations so widely used today in wireless
telegraphy, for instance, those produced by Duddell's arc or by the
three-electrodes lamps used as oscillators. The production and the maintenance
of oscillations in all these systems mainly result from the presence, in the
oscillating circuit, of something analogous to a negative resistance\footnote{The concept of ``negative resistance'' was introduced by the German physicist Hans \cite[p. 568]{Luggin1888}.}.''

\end{quote}

At the end of June 1920, the French engineer Jean-Baptiste Pomey (1861-1943) made use of a cubic function to represent the tension-current characteristic ($v,i$) of the \textit{singing arc} \cite[p. 380]{Pomey1920}. Three weeks later, \cite[p. 703]{VdP1920} employed the same type of modeling for the triode. Ten years later, on march 11$^{th}$ 1930, during a conference given at the \'{E}cole Sup\'{e}rieure d'\'{E}lectricit\'{e} (today Sup'Elec), \cite[p. 300]{VdP1930} claimed that the ``nearly sinusoidal'' oscillations produced by the \textit{singing arc} represent one of the ``very nice example of relaxation oscillations''.\\

Thus, in the early 1930s, at least three analogous devices, namely the \textit{series dynamo machine}, the \textit{singing arc} and the \textit{triode}, with nonlinear tension-current characteristic ($v,i$) were known. Let's focus on the main properties of the \textit{singing arc}.

\subsection{Arc characteristics}

According to the British engineer Mrs Hertha Ayrton (1854-1923) the arc ``static characteristic'' (static because the voltage and currents are supposed to increase or decrease slowly during the experiments as it is the case when the oscillatory circuit $LC$ is disconnected.), i.e. the tension-current characteristic ($v,i$) of the \textit{singing arc} may be represented by an hyperbola \cite{Ayrton1895}:

\begin{equation}
e = a + \frac{b}{i}
\end{equation}

\vspace{0.1in}

where $a$ and $b$ are constants depending on the conditions of the experiments (supply current, length of the arc, thermal conductivity and nature of the electrodes, as well as on the temperature and pressure of the surrounding gas...).
The French physicist Henri Bouasse (1866-1953) plotted some arc ``static characteristic'' ($DiH$) and ($ASC$) as exemplified on Fig. \ref{fig6}:

\begin{figure}[htbp]
\centerline{\includegraphics[width=7.44cm,height=7cm]{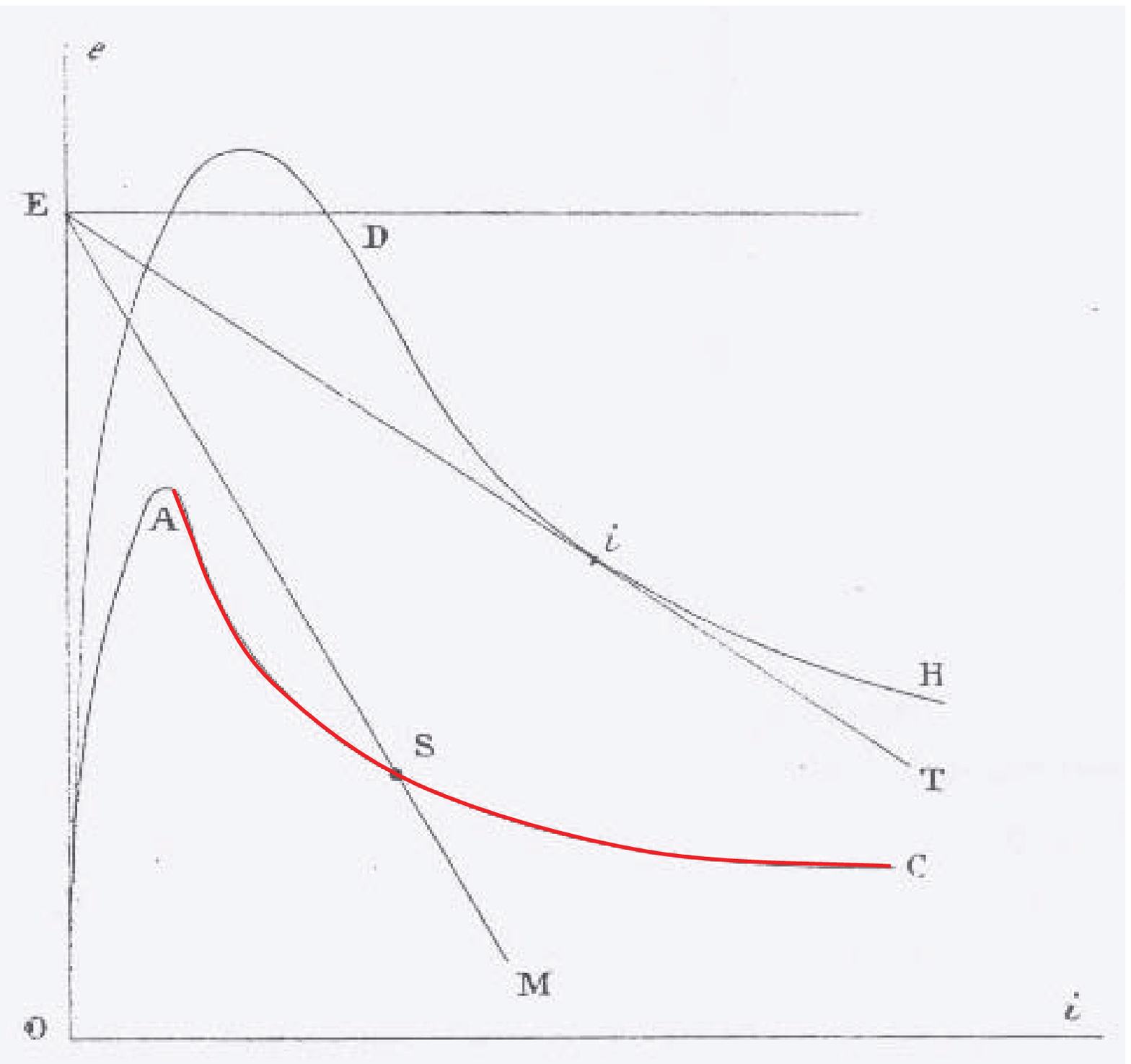}}
\caption{Singing arc static characteristic, \cite[p. 250]{Bou1924}}
\label{fig6}
\end{figure}

As noticed by Bouasse, the curve must not be extended to the asymptote $i = 0$ since for small currents the above relation (6) is no longer valid (between $O$ and $A$ in Fig. \ref{fig6}).

\subsection{Arc hysteresis}

\begin{figure}[htbp]
\centerline{\includegraphics[width=11.08cm,height=13.79cm]{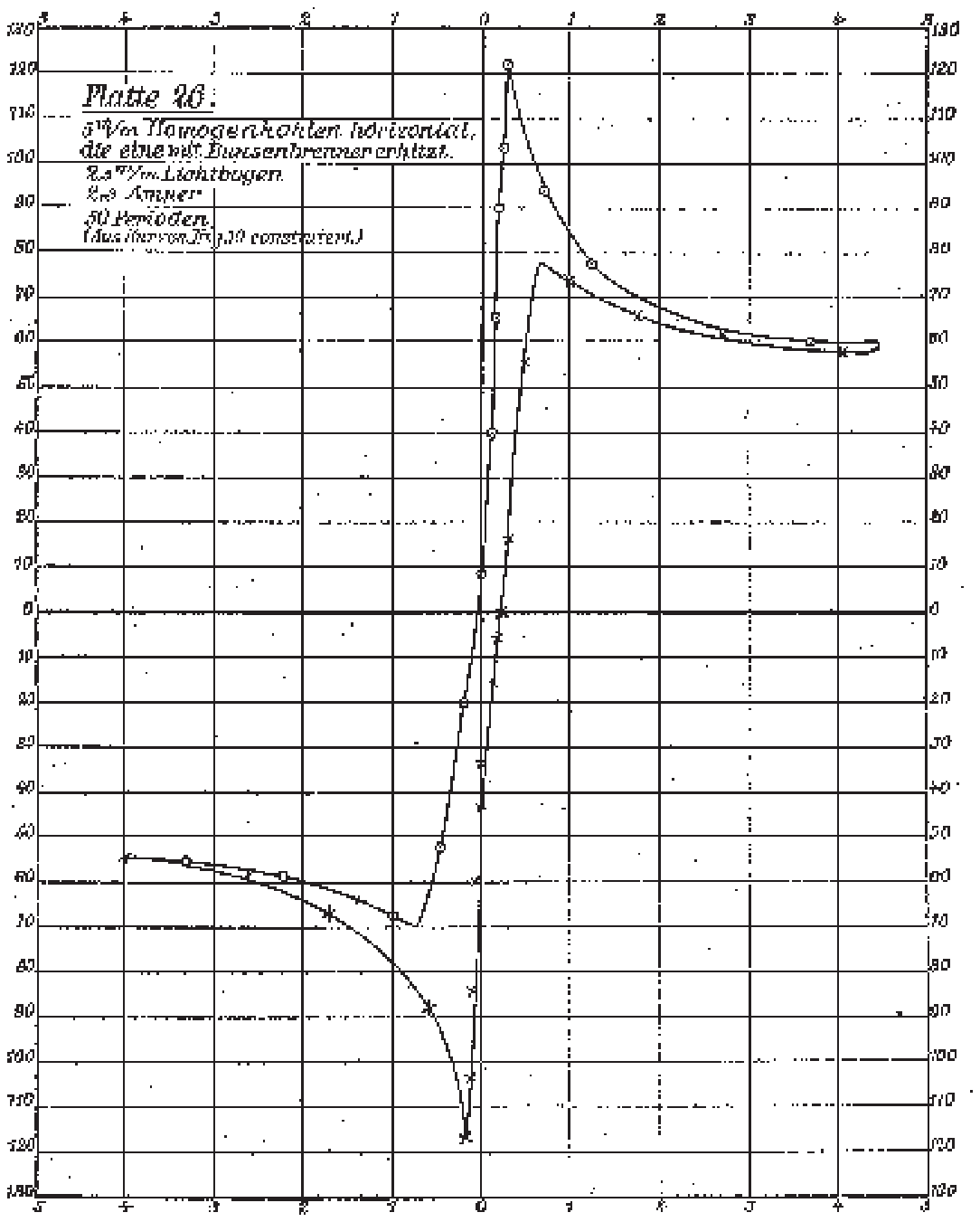}}
\caption{Arc hysteresis, arc ``dynamic characteristic'', \cite[p. 305]{Simon1905}}
\label{fig7}
\end{figure}

In the case of \textit{Duddell's arc}, since an oscillatory circuit $LC$ has been connected to the arc, an alternating current flowing out of this oscillatory circuit superposes itself on the direct current in the arc, and thus turns it into an oscillating current, which lags 180 degrees behind the current in the shunt circuit, i.e., it is increasing while the shunt current is decreasing, and \textit{vice versa}. In 1905, the German physicist Hermann Theodor Simon (1870-1918) highlighted this effect that he called ``Lichtbogenhysteresis'', i.e. ``arc hysteresis'' \cite[p. 305]{Simon1905}. When the voltage and current are varied very rapidly (as it is the case when the oscillatory circuit $LC$ is connected), the curve is called ``dynamic characteristic'' and the rising voltage curve is different from the falling one (See Fig. \ref{fig7}).

\begin{figure}[htbp]
\centerline{\includegraphics[width=5.93cm,height=7cm]{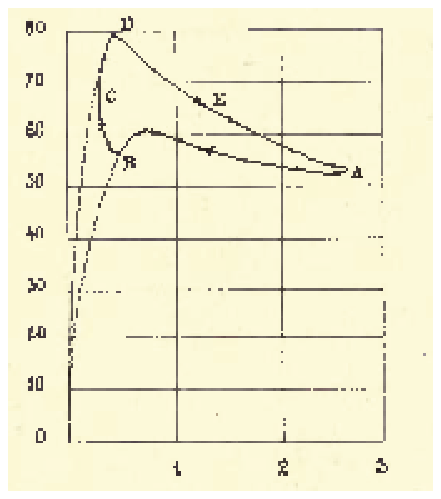}}
\caption{Arc ``dynamic characteristic'', \cite[p. 150]{Ruhmer1908}}
\label{fig8}
\end{figure}

\begin{figure}[htbp]
\centerline{\includegraphics[width=6.95cm,height=7cm]{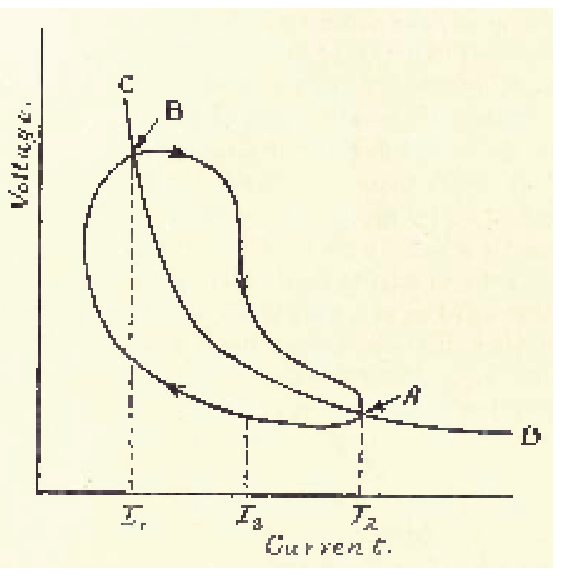}}
\caption{Arc ``dynamic characteristic'', Arc ``static characteristic'' (CD), \cite[p. 115]{Coursey1919}}
\label{fig9}
\end{figure}

So, when the arc is functioning its ``static characteristic'' is no longer valid and must be replaced by its ``dynamic characteristic''. Figure \ref{fig8} represents a ``zoom'' of the right top of the Fig. \ref{fig7} and illustrates with Fig. \ref{fig9} diagrammatically one form taken by the above curve under these conditions, i.e., when the oscillatory circuit $LC$ is connected. Let's notice that in this case the ``dynamic characteristic'' becomes a closed curve. Since both $v(t) >0$ and $i(t) >0$ in the dynamic regime, the hysteresis curve in this oscillation did not pass through the origin. This hysteresis loop is therefore not ``pinched'' \cite{Chua2011} as would be the case if either $v(t)$ or $i(t)$ attains the ``zero'' value during each period.

This phenomenon is closely related to the fact that the number of ions existing between the electrodes depends upon the current and the
temperature of the electrodes at the preceding instant, the number increasing as the current and electrode temperature increase. More precisely, ``arc hysteresis'' transcribes the fact that the temperature is lagging behind the current. Hence with rising current the number of ions is less at a given current value than with decreasing current at the same current value, and the voltage necessary to produce a given current is greater in the first case than in the second for the same reason. Let's consider the point A of Fig. \ref{fig9}, with a large current through the arc, the electrodes will be at a high temperature, and consequently there will be a great supply of ions to maintain the arc. If now the current is suddenly diminished from its value $I_2$ at this point towards some other lower value, say $I_3$, the temperature of the electrodes, and therefore also the supply of ions, will not \textit{immediately} decrease to the values they would normally have for the current in question, since the electrodes will require a certain time
interval for cooling down. The arc gap will therefore be in a more conducting state than usual, for the current $I_3$, and therefore the P.D. between its terminals, will be lower. Now, lets' consider the point B of Fig. \ref{fig9} with a small current $I_1$ and an impoverished supply of ions. When the current is increased the electrode temperatures do not immediately respond to the change, so that consequent upon this the gap is less ionised than usual, and exhibits therefore a higher apparent resistance or what is the same thing, the P.D. between its terminals is higher than the usual value for the same current. So, we can say that if, the current in the circuit is alternately increased and decreased between two values corresponding
to points A and B of the curve, the voltage will vary in opposite manner, that is, respectively decrease and increase. In other words, the ratio $dV/dI$ of corresponding voltage and current variations is negative, which is often expressed in saying that the alternating-current resistance of the arc is negative (since the resistance $R$ of a conductor is precisely equal to the ratio $dV/dI$ of corresponding voltage and current variations).

Moreover, in a note at the French \textit{Comptes Rendus} of the Academy of Sciences in which he established the second-order differential equation of the singing arc oscillations\footnote{It has been stated by one of us \cite{GiBlondel} that \cite[p. 380]{Pomey1920} had already stated this equation in 1920 by using a cubic function in order to represent the characteristic of the singing arc.}, \cite[p. 900]{Blo1926} defined thus the ``arc hysteresis'' phenomenon:\\

\begin{quote}
``Ce mot, bien que form\'{e} tr\`{e}s correctement au point de vue \'{e}tymologique, est devenu impropre pour l'arc par le fait qu'il est déj\`{a} employ\'{e} en magn\'{e}tisme pour d\'{e}finir un ph\'{e}nom\`{e}ne de retard sensiblement ind\'{e}pendant de la fr\'{e}quence, tandis qu'ici le retard va en d\'{e}croissant rapidement quand la fr\'{e}quence augmente.\footnote{``Although this word is correctly formed on an etymological point of view, it has became inappropriate for the arc because it has been already used in magnetism for defining a phenomenon of delay strictly independent of the frequency, while in this case the delay is decreasing quickly as the frequency increases.''}''
\end{quote}

\section{Memristor: the missing circuit element}

As previously recalled, starting from the logical as well as axiomatic points of view, \cite{Chua1971} was led to postulate the existence of a fourth circuit element to which he gave the name \textit{memristor} since it behaves like a nonlinear resistor with memory. By considering the three classical circuit elements, the \textit{capacitor}, the \textit{resistor}, the \textit{inductor} and the ``relationships'' linking the variables involved, respectively: ``Volta's law'' $q = Cv$ (Eq. (1)), ``Ohm's law'' $v = Ri$ (Eq. (2)) and ``Faraday's law'' $\varphi = Li$ (Eq. (3)), he deduced, for the sake of \textit{completeness}, that only one ``relationship'', i.e. only one law, remained undefined.

\subsection{Theoretical imperfections}

The philosopher of sciences Dudley Shapere \cite{Shapere1974} distinguished three kinds of scientific problems, first, problems related to scientific domain concerning for example the clarification and extension of the field, then the theoretical problems that require that one accounts, in a more profound manner, for a certain theory of the domain, and finally the \textit{theoretical imperfections}. These, arise eventually when the theory was developed in response to a given theoretical problem. The theory can be revealed as incomplete with respect to the considered domain, be a simplification, or cause problems that Shapere called \textit{incompleteness of the black box}. In fact, these three types of theoretical imperfections are distinguished from each other starting from the domain considered by the theory and the observed correspondence between theory and reality. Thus, if we consider a scientific domain as a whole corpus of information that a theory must account, we can say that a theory can be revealed as incomplete with respect to the considered domain if, on one hand there is no reason to suppose that the theory is not fundamentally correct and, on the other hand there is no reason to suppose that this theory can not be completed so that it can account for an unexplained part of the corpus.\\

Following this idea of \textit{completeness}, \cite{Chua1971} postulated the existence of a fourth circuit element in addition to the three others that he called it \textit{memristor}. Then, he established the ``relationship'' between magnetic flux $\varphi$ and charge $q$:

\begin{equation}
\varphi = Mq
\end{equation}

\vspace{0.1in}
 
\begin{figure}[htbp]
\centerline{\includegraphics[width=7.56cm,height=7.34cm]{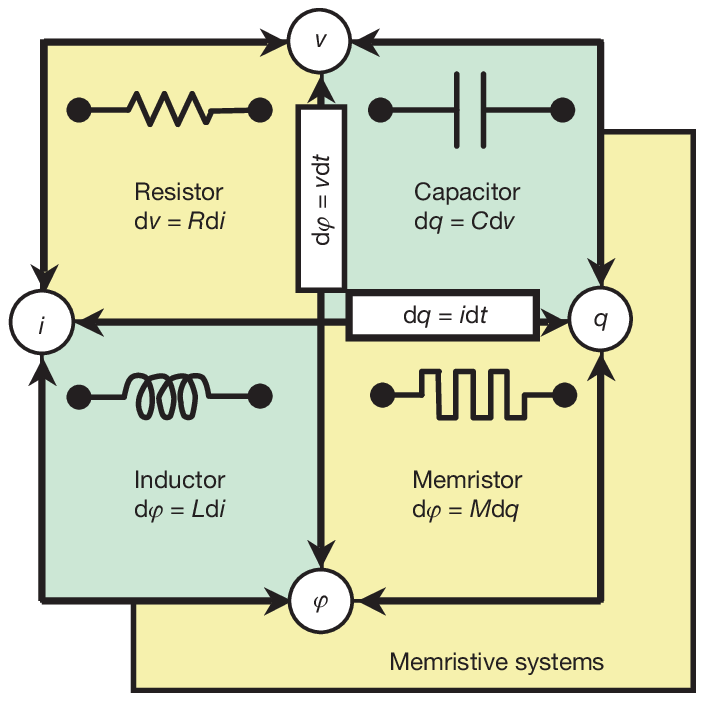}}
\caption{The four fundamental two-terminal circuit elements, \cite[p. 80]{Strukhov2008}}
\label{figMem}
\end{figure}

We suggest to call this relationship: Chua's law. A that time, he also proposed a symbol (See Fig. \ref{figMem}).

\subsection{Memristor DC Characteristics}

In order to emphasize a strong analogy between the memristor properties and and that of the singing arc let's focus on the interesting properties 3-5 established by Prof. Chua in his second publication on this subject \cite[p. 211 and next]{Chua1976}. Let's begin with the:\\

\begin{quote}
``\textit{Property 3--DC Characteristics}\\
A time-invariant current-controlled memristive one-port under dc operation is equivalent to a time-invariant current-controlled
nonlinear resistor if $f(x, I) = 0$ has a unique solution $x = X(I)$ such that for each value of $I \in \mathbb{R}$, the equilibrium
point $x = X(I)$ is globally asymptotically stable.''
\end{quote}

\vspace{0.1in}

\begin{figure}[htbp]
\centerline{\includegraphics[width=7.94cm,height=7cm]{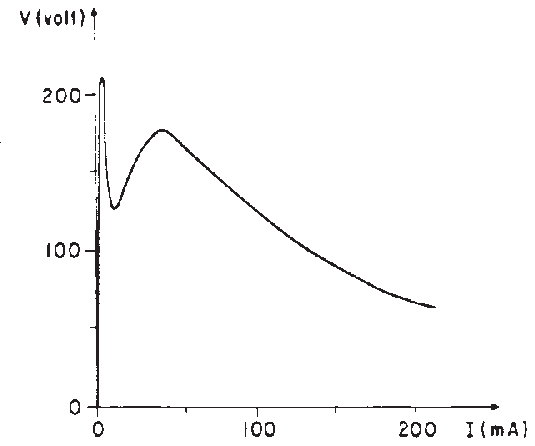}}
\caption{The dc characteristics of a short neon tube, \cite[p. 211]{Chua1976}}
\label{fig10}
\end{figure}

\cite[p. 211]{Chua1976} then observed that a time invariant memristive one-port under dc operation behaves just like a nonlinear resistor and took as an example the dc characteristics of a short neon tube (Fig. \ref{fig10}).

Starting from 25 mA Fig. \ref{fig6} resembles to Fig. \ref{fig10}. However, such similarity as been already pointed out by \cite[p. 992]{VdP1926d} who stated that the neon tube was another example of relaxation oscillator as well as the \textit{singing arc} and the \textit{triode}.

\begin{figure}[htbp]
  \begin{center}
      \includegraphics[width=7.44cm,height=7cm]{Fig6bis.eps}\\
      \includegraphics[width=7.54cm,height=7cm]{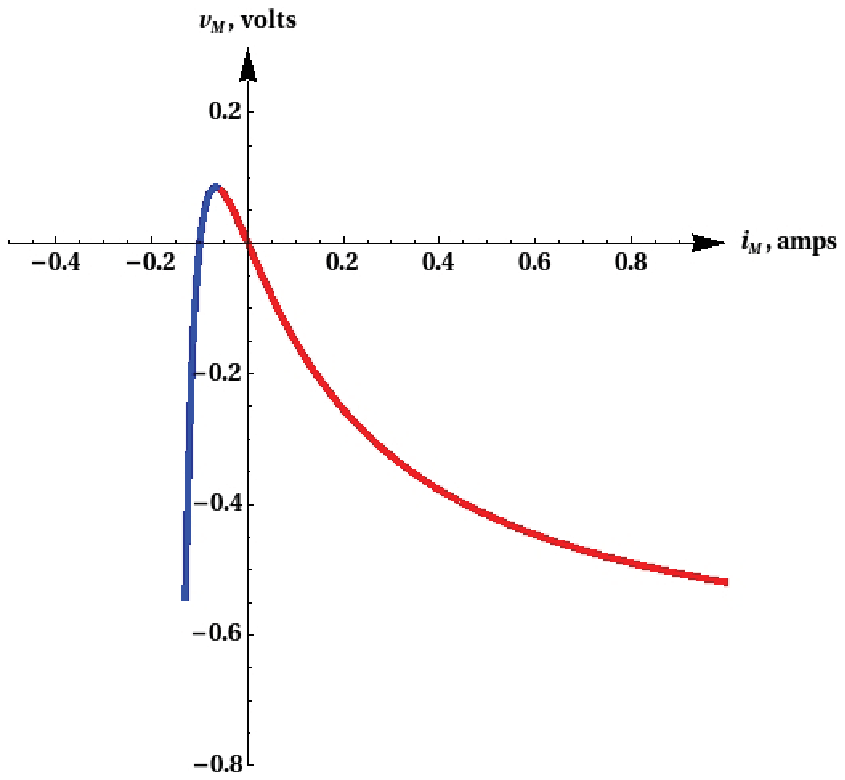} \\
      (a) Singing arc, \cite[p. 250]{Bou1924}\\
      (b) Memristor, \cite[p. 1574]{Muthuswamy2010} \\[0.2cm]
    \caption{The DC $v_M-i_M$ curve for the singing arc \& for the memristor.}
    \label{fig11}
  \end{center}
  \vspace{-0.5cm}
\end{figure}

If we compare now the singing arc static characteristic Fig. \ref{fig6} with the DC $v_M-i_M$ curve of the memristor (Fig. \ref{fig11}) plotted in a recent publication of \cite[p. 1574]{Muthuswamy2010} the comparison between both curves shape is obvious. Especially if you take into account the legend which specifies that the locally-active region highlighted in red on Fig. \ref{fig11} begins far from the origin exactly like in the case of the singing arc (Cf. supra).

But, we can go further by comparing the equation of each DC $v_M-i_M$ curve. For the singing arc it has been given by Eq. (5). In their paper \cite[p. 1580]{Muthuswamy2010} gave for the memristor the following equation:

\begin{equation}
v_M =i_M\left( -1 + \frac{i_M^2}{\left( i_M + \alpha \right)^2}\right)\beta
\end{equation}

\vspace{0.1in}

This Eq. (7) may be written as:

\[
v_M =i_M\left( -1 + \frac{1}{\left( 1 + \frac{\alpha}{i_M} \right)^2}\right)\beta
\]

\vspace{0.1in}

A Taylor series expansion of the ratio up to order three reads provided that $\alpha/i_M \ll 1$:

\[
\frac{1}{\left( 1 + \frac{\alpha}{i_M} \right)^2} = 1 - \frac{2\alpha}{i_M} + \frac{12\alpha^2}{i_M^2} + h.o.t.
\]

\vspace{0.1in}

Thus, we find for the DC $v_M-i_M$ curve of the memristor:

\begin{equation}
v_M = -2\alpha \beta + \frac{12\alpha^2}{i_M} + \ldots
\end{equation}

\vspace{0.1in}

Since Eq. (5) \& Eq. (8) are identical, we can conclude that the singing arc ``static characteristic'' corresponds exactly to Chua's DC $v_M-i_M$ curve of the memristor.\\

\subsection{Memristor: an hysteretic system}

According to \cite[p.209]{Chua1976} memristive systems are hysteretic. In order to study this effect let's focus on the\\

\begin{quote}
``\textit{Property 4--Double-Valued Lissajous Figure Property}
A current-controlled memristive one-port under periodic operation with $i (t) = I cos \omega t$ always gives rise to a $v-i$ Lissajous figure whose voltage $u$ is at most a double-valued function of $i$
\end{quote}

\vspace{0.1in}

\begin{figure}[htbp]
\centerline{\includegraphics[width=7.46cm,height=7cm]{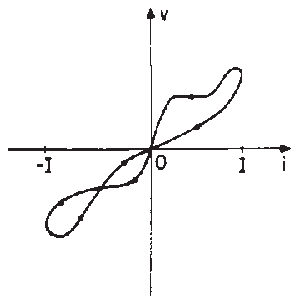}}
\caption{Hysteretic effect, \cite[p. 212]{Chua1976}}
\label{fig12}
\end{figure}

Precisely, it has been stated for the singing arc ``dynamic characteristic'' that the rising voltage curve is different from the falling one. By plotting any parallel to the $v-axis$ it is easy to check on Fig. \ref{fig7}-\ref{fig9} that such a curve is double-valued. Moreover, the closed curve used by \cite[p. 212]{Chua1976} to emphasize this property and plotted on Fig. \ref{fig12} is analogous to the ``dynamic characteristic'' presented on Fig. \ref{fig7}.

\textit{Property 5} was summarized by \cite[p. 2132]{Chua1976} by saying that ``the $v-i$ curve is odd symmetric with respect to the origin.'' It is the case of Fig. \ref{fig6} if we model it with a cubic function.\\

Then, let's compare this sentence with Chua's conclusion \cite[p. 220]{Chua1976}\\

\begin{quote}
``Various generic properties of memristive systems have been derived and shown to coincide with those possessed by many physical devices and
systems. Among the various properties of memristive systems, the frequency response of the Lissajous figure is especially
interesting. As the excitation frequency increases toward infinity, the Lissajous figure shrinks and tends to a straight line passing through the origin-except for some pathological cases where the bibs\footnote{bounded-input bounded-state.} stability property is not satisfied. The physical interpretation of this phenomenon is that the system possesses certain inertia and cannot respond as rapidly as the fast variation in the excitation waveform and therefore must settle to some equilibrium state. \textbf{This implies that the hysteretic effect of the memristive system decreases as
the frequency increases} and hence it eventually degenerates into a purely \textit{resistive} system.''
\end{quote}

\vspace{0.1in}

and that of \cite[p. 900]{Blo1926} in which he claimed that ``hysteresis effect is decreasing quickly as the frequency increases''.\\

At last, let's notice that the Duddell's singing arc circuit in Fig. \ref{fig5} can be proved via Circuit theory principles, to be equivalent to the simplest chaotic circuit as exemplified in \cite{Muthuswamy2010} paper. Since one more state variable is needed, in addition to $L$ and $C$ in Duddell's circuit and since one nonlinearity is also needed to obtain chaos as in \cite{Muthuswamy2010} article, and since Duddell's circuit exhibits ``hissing''sounds, and, ``noise-like'' chaotic waveforms as observed with an oscilloscope by \cite{Blo1905}, it follows that the Duddell's singing arc circuit in Fig. \ref{fig5} is experimentally a chaotic circuit, and hence the singing arc must be a memristor.

\section{Discussion}
\label{conc}
In this work, we have first recalled the history of the three classical circuit elements namely \textit{capacitor}, \textit{resistor}, \textit{inductor} and the relationships, i.e. \textit{Volta's law}, \textit{Ohm's law} and \textit{Faraday's law} involving the four fundamental circuit variables: the \textit{current}, the \textit{voltage}, the \textit{charge} and the \textit{magnetic flux} in order to highlight the \textit{incompleteness} of the Circuit Theory which lead Professor Chua to postulate the existence of a fourth element he called \textit{memristor}.
Then, we focused on the properties of an old nonlinear device called \textit{singing arc}, originally used in Wireless Telegraphy before the invention of the triode. By comparing the \textit{singing arc} to the \textit{memristor} we have highlighted an analogy between some of their properties. The \textit{memristor} DC characteristic curve has exactly the same hyperbolic profile and the same equation as the \textit{singing arc} ``static characteristic''. Moreover, both \textit{memristor} and \textit{singing arc} are hysteretic systems exhibiting the same kind of closed curve in order to describe the ``hysteresis effect''. So, we conclude that the \textit{singing arc} could be considered as the oldest \textit{memristor}. We believe that the hysteresis characteristic in Fig. \ref{fig7} should have intersected each other at the origin, forming a ``pinched'' hysteresisloop. We therefore conjectured that the slight discrepancy of the ``almost pinched'' characteristic near the origin is due on the one hand to the imperfection of instruments used in this measurement and, on the other hand to the great variability of the conditions of the experiments.

\section{Acknowledgments}

Authors would like to thank Dr. Bharathwaj Muthuswamy for his helpful advices.

\end{document}